\documentclass[10pt,twocolumn,twoside]{IEEEtran}

\usepackage[cmex10]{amsmath}
\usepackage{amssymb,amsfonts,colortbl}
\usepackage{cite}
\usepackage{graphicx}

\def \EE {{\mathbb{E}}}

\newcommand{\Prob}{\mathbb{P}}
\newcommand{\A}{\mathcal{A}}

\def\Esp#1{\mathbb{E}\left[#1\right]}

\def\fun#1{\mathrm{#1}}

\newcommand{\Tr}[1]{\mathsf{tr} \left( #1 \right) }
\newcommand{\Covar}[2]{ \mathsf{Cov}\left[ #1, #2 \right] }
\newcommand{\E}{\mathcal{E}}

\newcommand{\Rd}{R_d}
\newcommand{\Lf}{\mathcal{L}}
\newcommand{\Lfp}[1]{\Lf\left(#1\right)}

\newcommand{\Lfsp}[1]{\Lf^* \left( #1 \right) }

\newcommand{\Mq}[1]{M\!\left(#1\right)}
\newcommand{\Mqt}{\tilde{M}}

\newcommand{\At}{\tilde{\A}}

\newcommand{\F}{\mathcal{F}}

\newcommand{\fracpow}[3] {\left( \frac{#1}{#2} \right) ^{#3}}

\newcommand{\MP}{\mathcal{P}}

\newcommand {\GamF} [3] {#1_{\Gamma_{#2}, \Gamma_{#3} } }

\newcommand{\Normal}[2]{\mathcal{N} _{ #1,#2 } }
\newcommand{\Real}[1]{\mathfrak{Re}\left\{ #1 \right\}}

 \newcommand{\Comment}[1]{#1}

\newcommand{\R}{\mathbb R}

\newcommand{\N}{{\mathbb N}}

\newcommand{\comment}[1]{}
\newcommand{\X}{\underline{X}_N}
\newcommand{\Y}{\underline{Y}_N}
\newcommand{\ZZ}{\underline{Z}_N}
\newcommand{\Xv}{\underline{X}}
\newcommand{\Eqref}[1]{Eq.~(\ref{#1})}

\graphicspath{
{Figures/}
}

\begin{document}
\title{Matrix product representation and synthesis for random vectors: Insight from statistical physics\footnote{J.-Y. Tourneret and N. Dobigeon are gratefully acknowledged for fruitful discussions.
Work partially supported by the \emph{Young Research Team} Del Duca French Academy of Sciences Award, 2007. }}
\author{Florian Angeletti*, Eric Bertin, Patrice Abry,\\
\makebox{} \\
Universit\'e de Lyon, Laboratoire de Physique,
Ecole Normale Sup\'erieure de Lyon, CNRS\\
46 all\'ee d'Italie, F-69007 Lyon, France\\
{\tt ens-lyon.fr/PHYSIQUE/}, {\tt firstname.lastname@ens-lyon.fr}\\
}

\maketitle
\begin{abstract}
Inspired from modern out-of-equilibrium statistical physics models, a matrix product based framework permits the formal definition of random vectors (and random time series) whose desired  joint distributions are a priori prescribed. 
Its key feature consists of preserving the writing of the joint distribution as the \emph{simple} product structure it has under independence, while inputing controlled dependencies amongst components: This is obtained by replacing the product of distributions by a product of matrices of distributions.
The statistical properties stemming from this construction are studied theoretically: The landscape of the attainable dependence structure is thoroughly depicted and a stationarity condition for time series is notably obtained. 
The remapping of this framework onto that of Hidden Markov Models enables us to devise an efficient and accurate practical synthesis procedure. 
A design procedure is also described permitting the tuning of model parameters to attain targeted properties.
Pedagogical well-chosen examples of times series and multivariate vectors aim at illustrating the power and versatility of the proposed approach and at showing how targeted statistical properties can be actually prescribed.
\end{abstract}


 \begin{center} \bfseries EDICS Category: SSP-SNMD SSP-NGAU SSP-NSSP \end{center}

\section{Introduction}
\label{sec:intro}

In modern signal processing, it is very often needed that synthetic data are produced numerically in fast and efficient manners with (some of) their statistical properties being a priori prescribed as closely as desired from selected targets (marginal distributions, covariance, spectrum, multivariate distribution,\ldots). 
This is for instance the case when the performance of newly developed statistical analysis need to be assessed.
The theoretical derivation of such performance may turn too difficult to achieve, specially when it is intended to apply such tools to real-word data, whose properties are not well known. 
Instead, one can resort to Monte Carlo simulations: Performance are derived from averages over independent realizations of synthetic data, that are designed to resemble as closely as possible the real-world data. 
Another example stems from Bayesian estimation procedures, that generally involve Monte Carlo Markov Chain \cite{Berg2004} or variational based resolution schemes (cf. e.g., \cite{tlg2008,Bishop2006}) to generate numerically independent copies of hyper-parameters drawn from (possibly complicated) distributions, derived from the Bayesian formalism. 

For the numerical synthesis of Gaussian time series, with prescribed covariance function, the so-called \emph{Circulant Embeded Matrix} synthesis procedure \cite{DietrichNewsam1997, DemboMallowsShepp1989,DaviesHarte1987} is nowadays considered as the state-of-the-art solution and is currently widely used. 
It has then been extended to the synthesis of multivariate Gaussian time series, with prescribed auto- and cross-covariance, notably in \cite{WoodChan94,Chan99} (see also \cite{hpa10b} for variations).
The synthesis of non-Gaussian time series turns far more complicated as the prescription of the full joint distributions turns very difficult to achieve while that of the sole marginal distributions and covariance functions does not uniquely define the process. 
Several approaches were proposed to address such issues (cf. e.g., \cite{GrigoriuBook,hpa10} and references therein for reviews). 
Other strategies are based on surrogate techniques: Starting from the original real-world data of interest, surrogate copies are obtained by randomizing one attribute of the data (e.g., the phase of its Fourier transform) while maintaining another fixed (e.g., the amplitude of the Fourier transform) (cf. e.g., \cite{SchreiberSchmitz2000}). 
Recently, an \emph{optimal transport} procedure was proposed aiming at iteratively modifying the joint distribution of random vectors or time series to attain a given target \cite{BORGNAT:2012:A}. 
The general framework of Markov Chain simulation offers an alternative and broad class of solutions, focusing on the modelling of local dynamical properties, while not explicitly putting the emphasis on a direct prescription of the joint distributions of the process. 
Markov Chain schemes are also widely used to synthesize independent realizations of random vectors with prescribed properties (cf. e.g.,  \cite{Chen2000} for a review). 

The present contribution takes place in this long tradition of synthetic data design in signal processing, but is however rooted in a very different scientific field, that of statistical physics. 
Indeed, inspired from the exact solutions of stochastic out-of-equilibrium models describing particle diffusion on a one-dimensional lattice with volume exclusion, like the Asymmetric Simple Exclusion Processes (ASEP) \cite{DerridaASEP93,Mallick97,Evans2007}, the design procedure proposed here is founded on the request that the joint distribution of the random vector or random times series $\underline{X}_N={x_1, x_2, \ldots, x_N} $ must be written as a product $ P_{\underline{X}} \propto \prod_{k=1}^{N} R_d(x_k)$, as in the case of independent components.
However, the $R_d$s are no longer univariate distributions, but instead $d$ dimensional matrices of valid unnormalized distributions.
In statistical physics, this matrix product form of the joint distribution was envisaged as a practical ansatz to find exact solutions of the master equation associated e.g. to the ASEP model.
In signal processing, this enables us to devise a theoretically powerful and practically versatile and efficient procedure to define and generate numerically random vectors or random times series, with a priori prescribed statistical properties. 
A preliminary and partial attempt relying on specific choices of matrices and focusing on stationary time series only was presented in \cite{Angeletti2012:ICASSP}. 
It is here extended to a more general and versatile setting.

Definitions of this matrix product joint distribution framework and the general consequences in terms of statistical properties are derived in Section \ref{sec:theory}.
Notably, a general condition to ensure stationarity of the time series is obtained. 
In Section~\ref{sec:asympt}, the dependence structures that can be achieved from this model are studied in-depth and the practical tuning of the time-scales that can be input in these dependence structures is established in details.
In Section~\ref{sec:synth}, it is first shown how this matrix product framework can be explicitly recast into that of Hidden Markov Models, hence providing us with a fast and efficient practical synthesis procedure; and it is, second, explained how targeted marginal distributions and other statistical properties can be attained. 
Section~\ref{sec:illust} consists of two different sets of pedagogical examples aiming at illustrating the efficiency and versatility of the tool: First, sets of time series whose marginal distributions, covariance functions and some higher order statistics are jointly prescribed, are produced; Second, sets of random vectors, with different marginal distributions are synthesized and it is shown how to tune the correlation and fourth order statistics amongst the components of such vectors while maintaining fixed the marginal distributions. 

\section{Joint probability as matrix product}
\label{sec:theory}

\subsection{Definitions and general setting}

Inspired from out-of-equilibrium statistical physics models (cf. e.g. \cite{DerridaASEP93,Mallick97,Evans2007}), a general framework for the design of joint distribution functions for random vectors $\underline{X}_N  $, of size $ N$ is devised.  
It is based on product of matrices and relies on $4$ different key ingredients defined below. 

First, let $d \in \N^*$. 
Second, let $\A$ denote a fixed non-zero non-random $d \times d$ matrix, with positive entries, and $\Lf(M)$ an associated linear form defined (for any matrix $M$) as: 
\begin{equation} 
\Lf(M)= \Tr {\A^T M}. 
\end{equation}
Third, let  $ \E $ denote a fixed non-zero non-random $d \times d$ matrix, with positive entries $ \E_{i,j}  $. 
For reasons made clear in Section~\ref{sec:asympt}, $ \E $ will be here referred to as the \emph{structure matrix}.
Fourth, let Matrix $\MP$ consists of a  $d \times d$ set of valid (normalized) distribution functions $\{ \MP_{i,j}(x) \}_{i=1,\ldots,d ; j=1,\ldots,d}$. 
\begin{equation}
\label{equ-Rd} 
\Rd(x)=\E \otimes \MP(x),
\end{equation}
where $\otimes$ denotes entry-wise matrix multiplication.
Let $\Xv \equiv \{X_n\}_{1\le n \le N}$ denote the random vector explicitly defined via its joint distribution: 
\begin{equation} 
\label{eqn:prob}
\Prob_{\underline{X}}(x_1,\dots,x_N)=\frac{\Lfp { \prod_{k=1}^N \Rd(x_k) } }{\Lfp {\E^N} },  
\end{equation}
where $\prod_{k=1}^N \Rd(x_k)=R_d(x_1) \ldots R_d(x_N)$ denotes the oriented product (i.e., the order of the factors is fixed).
The joint choice of positive entries for matrices $\A$ and $\E$, and of valid distribution functions in $\MP $ is sufficient to ensure that Eq. (\ref{eqn:prob}) defines a valid joint distribution function.

\subsection{Marginal distributions, moments and dependence}

From these definitions, using commutativity of integration and matrix product, a number of statistical properties of $\underline{X}_N  $ can be analytically derived.

Univariate or marginal distributions take explicit forms:
\begin{equation}
\label{eqn:pdf:univ}
\begin{aligned}
\Prob_k(X_k=x_k) &= \frac{\int \Lfp{\prod_{i=1}^N R_d(x_i)} \prod_{i\ne k} dx_i}{\Lfp{\E^N}} \\
&= \frac{\Lfp{ \int R_d(x_1) dx_1 \dots R_d(x_k) \dots R_d(x_N) dx_N }}{\Lfp{\E^N}}\\
&= \frac{\Lfp{\E^{k-1}R_d(x_k) \E^{N-k}}}{\Lfp{\E^N}}.
\end{aligned}
\end{equation}
This shows that the marginal distributions necessarily consist of weighted linear combinations of the $\MP_{i,j}$, 
\begin{equation}
\label{eq:inv:univ}
\Prob_k(X_k=x) = \sum_{i,j} c_{i,j,k} \MP_{i,j}(x),
\end{equation}
where the $c_{i,j,k}$ are functions of $\A $ and $\E$, satisfying, for all $k$, $\sum_{i,j} c_{i,j,k}=1$.

Further, let us define the collection of matrices, $\Mq{q}  = \int_\R x^q \Rd(x) dx= \E \otimes \int_\R x^q \MP(x) dx $, whose entries read:
\begin{equation}
\label{eqn:Mq}
\Mq{q}_{i,j} = \E_{i,j} \int_\R x^q \MP_{i,j}(x) dx.
\end{equation}
Univariate moments also take closed-form expressions: 
\begin{equation} 
\label{eqn:moments:univ}
\Esp{X_k^q} =  \frac{\Lf ( \E^{k-1}\Mq{q}\E^{N-k})}{\Lf(\E^N)}.
\end{equation}
The $p$-variate distributions (with  $k_1<\dots<k_p$) can also be made explicit,
\begin{equation}
\label{eqn:pdf:multi}
\begin{split}
&\Prob(X_{k_1}=x_{k_1}, \dots, X_{k_p}=x_{k_p})= \\
& \frac{\Lfp{\E^{k_1-1} \left( \prod_{r=1}^{p-1} R_d(x_{k_r}) \E^{k_{r+1}-k_{r}-1}  \right) R_d(x_{k_p}) \E^{N-k_{p}} }}{\Lfp{\E^N}}
 \end{split},
\end{equation}
as well as the $p$-variate moments (with $q_r $ the order associated to entry $k_r$): 
\begin{equation} 
\label{eqn:moments:multi}
\begin{split}
&\Esp{\prod_{r=1}^p X_{k_r}^{q_r}}=\\
& \frac{\Lf \left(\E^{k_1-1} \left( \prod_{r=1}^{p-1} \Mq{q_r} \E^{k_{r+1}-k_{r}-1}  \right) \Mq{q_p} \E^{N-k_{p}} \right) }{\Lf(\E^N)}. 
 \end{split}
\end{equation}
\Eqref{eqn:moments:multi} constitutes a key result with respect to applications as it clearly shows that the joint statistics of order $q$ of $\X$ can be prescribed by the sole selection of matrices $M(q)$. 
This will be explicitly used in Section \ref{sec:illust}.

\subsection{Stationarity condition}
\label{sec:stationary}

\Eqref{eqn:prob} shows that the in general non-stationary nature of  $\underline{X}_N $ stems from the non-commutativity of the matrix product. 
To enforce stationarity in $\underline{X}_N $,  a commutativity property for $\A$ and $\E$ must be added, 
\begin{equation}
\label{eqn:static}
[\A^T,\E] \equiv \A^T \E - \E \A^T =0, 
\end{equation}
which ensures that $ \forall x, \Lfp{\E R_d(x) } = \Lfp{R_d(x) \E}$. 

Under \Eqref{eqn:static}, the marginal distribution (cf. \Eqref{eq:inv:univ}) simplify to: 
\begin{equation}
\label{eqn:static:updf}
\Prob_S(X=x)= \frac{\Lfp{ \Rd(x) \E^{N-1} } }{\Lfp{\E^N}} = \sum_{i,j} c_{i,j} \MP_{i,j}(x),
\end{equation}
and p-variate distributions and moments become (cf. \Eqref{eqn:pdf:multi} and \Eqref{eqn:moments:multi})
\begin{equation}
\label{eqn:static:pdf}
\begin{split}
&\Prob(X_{k_1}=x_{k_1}, \dots, X_{k_p}=x_{k_p})= \\
& \frac{\Lfp{\left( \prod_{r=1}^{p-1} R_d(x_{k_r}) \E^{k_{r+1}-k_{r}-1}  \right) R_d(x_{k_p}) \E^{N-(k_{p}-k_{1})-1} }}{\Lfp{\E^N}},
 \end{split}
\end{equation}
\begin{equation} 
\label{eqn:static:moments}
\begin{split}
&\Esp{\prod_{r=1}^p X_{k_r}^{q_r}}=\\
& \frac{\Lfp { \left( \prod_{r=1}^{p-1} \Mq{q_r} \E^{k_{r+1}-k_{r}-1}  \right) \Mq{q_p} \E^{N-(k_{p}-k_1)-1} } } {\Lfp{\E^N}}. 
 \end{split}
\end{equation}
These relations clearly indicate that the vector $\{ X_n \}_{1\le n \le N}$ can now be regarded as a stationary time series: All joint statistics depend only on time differences, $k_{r+1}-k_{r}$.

The sole matrix  $\A$ satisfying \Eqref{eqn:static} for all matrices $\E$, is the identity matrix, in which case  
$\Lf$ consists of the trace operator. 
However, the trace operator also induces automatically a circular correlation structure, $ l \ge 2 k$, $\EE X_k X_l = \EE X_k X_{N+2k-l}$,  a highly undesirable consequence for application purposes.

Alternatively, stationarity can be obtained by choosing jointly specific pairs $(\A, \E)$, such as 
\begin{equation}
\label{equ:Astat}
\A_{i,j}= (1/d)
\end{equation} 
and $\E$ a so-called \emph{doubly stochastic} matrix, defined as:
\begin{equation}
\label{equ:EDS}
\forall i,j ,\quad \sum_k \E_{k,j}= \sum_k \E_{i,k} \equiv 1 .
\end{equation} 
Indeed, such choices yield  
\[ \forall k,\, M, \quad \Lfp{M\E^k}= \Lfp{M}, \]
that leads to further simplifications of \Eqref{eqn:static:updf} to \Eqref{eqn:static:moments}.
The marginal and partial distributions of $\X$   no longer explicitly depend on the sample size $N$, also the marginal distribution is independent of $\E$: 
\begin{equation}
\label{eqn:sto:updf}
\Prob_k(X_{k}=x_k)= \Lfp{\Rd(x_k)},
\end{equation}
\begin{equation}
\label{eqn:sto:mpdf}
\begin{split}
\Prob(X_{k_1}=x_{k_1}, \dots, X_{k_p}=x_{k_p})= \\\Lfp{\left( \prod_{r=1}^{p-1} R_d(x_{k_r}) \E^{k_{r+1}-k_{r}-1}  \right) R_d(x_{k_p}) } ,
\end{split}
\end{equation}
\begin{equation}
\label{eqn:sto:moments}
\Esp{\prod_{r=1}^p X_{k_r}^{q_r}}=
 \Lfp { \left( \prod_{r=1}^{p-1} \Mq{q_r} \E^{k_{r+1}-k_{r}-1}  \right) \Mq{q_p} }.
 \end{equation}

Elaborating on a preliminary work (cf. \cite{Angeletti2012:ICASSP}), this particular setting will be used in  Section \ref{sec:static} to design efficiently stationary time series.

 \section{Dependence structure}
\label{sec:asympt}

Eqs.~(\ref{eqn:pdf:multi}) and (\ref{eqn:moments:multi}) indicate that Matrix $\E $ essentially controls the dependence structure within Vector $\underline{X}_N$, hence its name, via the collection of its powers $\E^n$, $n=1,\ldots, N$, while the matrices $\Mq{q}$ fix the amplitudes of the dependencies at order $q$. 
Analyzing the forms possibly taken by the $\E^n$ is hence crucial to understand the potential dependence structures of $\underline{X}_N$ achievable in this framework.
Notably, the classical distinction between diagonalizable and non-diagonalizable $\E$ plays a crucial role. 
This is studied in detail in this section for the correlation structure.

\subsection{Diagonalizable structure matrix}
\newcommand{\Eig}{E}

Let us assume that $\E$ is diagonalizable, with eigenvalues $\lambda_1,\dots,\lambda_r$.
Then, $\E$ can be decomposed into $r$ sub-matrices $\Eig_1,\dots,\Eig_r$ such that
\begin{equation}
\label{equ:subE}
\E^k= \sum_{i=1}^{r} \lambda_i^k \Eig_i, 
\end{equation}
and \Eqref{eqn:pdf:univ} can be rewritten as:
\begin{equation}
\label{eqn:margdiag}
\Prob_k(X_k=x)= \frac{\sum_{i,j}  \lambda_i^{k-1} \lambda_j^{N-k}  \Lfp{\Eig_i \Rd(x) \Eig_j}} {\sum_{i=1}^{r}  \lambda_i^N \Lfp{\Eig_i}}, 
\end{equation}
which explicits the dependence in $k$ and is reminiscent of \Eqref{eq:inv:univ} given that  the $ \Lfp{\Eig_i \Rd(x) \Eig_j}$ consist of linear combinations of the $\MP_{l,m}(x)$.
Further, for the 2-sample statistics, or covariance function, \Eqref{eqn:moments:multi} simplifies to: 
\begin{equation}
\label{equ:corrdiag}
\Esp{X_k X_l} =  \frac{\sum_{i,j,m}  \lambda_i^{k-1} \lambda_j^{N-l} \lambda_m^{k-l-1}  \Lfp{\Eig_i \Mq{1} \Eig_m \Mq{1} \Eig_j}} {\sum_{i=1}^{r}  \lambda_i^N \Lfp{\Eig_i}}. 
\end{equation}
Assuming stationarity, i.e., \Eqref{eqn:static}, the relation above further reduces to: 
\begin{equation}
\Esp{X_k X_l} =  \frac{\sum_{j,m}   \lambda_j^{N-2} \fracpow{\lambda_m}{\lambda_j}{k-l-1}  \Lfp{\Mq{1} \Eig_m \Mq{1} \Eig_j}} {\sum_{i=1}^{r}  \lambda_i^N \Lfp{\Eig_i}},
\end{equation}
which shows that the covariance function consists of the sum of weighted exponential functions $\exp(-(k-l-1) (\ln \lambda_j - \ln \lambda_m))$, with at most $r_M(r_M-1)/2= {r_M \choose 2}$ characteristic time scales, $ \tau_{j,m} = (\ln |\lambda_j| - \ln |\lambda_m|)^{-1}$, where $r_M$ stands for the number of  eigenvalues of Matrix $\E$ with different modulus.
A preliminary study of the covariance function in the stationary and diagonalizable case has been devised in \cite{Angeletti2012:ICASSP} and an example 
is worked out in Section~\ref{sec:circular}.

Without assuming stationarity, this exponential decrease of the covariance function still holds. 
Indeed, let us assume that $\Lfp{\Eig_1} \ne 0$ and that the norm of $\lambda_1$ is strictly larger than the norm of the other eigenvalues. 
Then, the normalisation term $\Lfp{\E^N}$ can be approximated in the limit $N\rightarrow +\infty$ as 
$ \Lfp{\E^N} \sim \lambda_1^N \Lfp{\Eig_1}$. 
Combining this asymptotic form with \Eqref{eqn:pdf:univ} yields:
 \begin{equation}
 \begin{aligned}
 \Prob_k(X_k= x) &= \sum_{i,j} \lambda_i^{k-1} \lambda_j^{N-k} \frac{\Lfp{\Eig_i \Rd(x) \Eig_j}}{ \Lfp{\E^N}}\\
 &\sim \frac{\Lfp{\Eig_1 R_d(x) \Eig_1}}{\lambda_1 \Lfp{\Eig_1}}, 
 \end{aligned}
 \end{equation} 
 and combining this result with \Eqref{equ:corrdiag} leads to: 
 \begin{equation}
  \begin{split}
 \Esp {X_k X_l} \sim  
  \sum_i \fracpow{\lambda_i}{\lambda_1}{k-l-1} \frac{\Lfp{ \Eig_1 \Mq{1} \Eig_i \Mq{1} \Eig_1 }}{\Lfp{\Eig_1} \lambda_1^2} 
 \end{split}
 \end{equation} 
These equations show that using a diagonalizable $\E$, with a dominant eigenvalue, implies that asymptotically, i.e., in the limit $N\rightarrow +\infty$, each component of Vector $\underline{X}_N$ shares the same univariate distribution, and that the autocovariance functions reads as a sum of exponential functions, depending only on  $|k-l|$. 
In the limit $N \rightarrow +\infty$, Vector $\underline{X}_N$ is hence asymptotically quasi-stationary.

\subsection{Non-diagonalizable structure matrices }

Non-diagonalizable matrices $\E$ draw a very different landscape. 
For illustration, let us consider the case where $\E= I_d + H$ with $I_d$ is the Identity matrix and H 	any nilpotent matrix of order $p+1$ (i.e., $H^{p+1} \equiv 0$ while $H^{k} \neq 0$, when $1\leq k \leq p$), chosen as a standard example of non diagonalizable matrix. 
Then, for any $k\ge p$, one has:
\[ \E^k= \sum_{j=0}^{p} {k \choose j } H^j,\]
which combined with \Eqref{eqn:pdf:univ} yields
\begin{equation}
\Prob_k(X_k=x_k)= \frac{\sum_{i,j\le p}  {k-1\choose i} {N-k \choose j} \Lfp{H^i \Rd(x) H^j}} {\sum_{i=1}^{p}  {N \choose i} \Lfp{H^i}}  
\end{equation}
and, for the covariance function:
\begin{equation}
\begin{aligned}
&\Esp{X_k X_l} = \\ 
&\frac{\sum_{i,j,m\le p}  {k-1\choose i} {k-l-1\choose m } {N-l \choose j} \Lfp{H^i \Mq{1} H^m \Mq{1} H^j}} {\sum_{i=1}^{p}  {N \choose i} \Lfp{H^i}}  
\end{aligned}
\end{equation}
To gain a better grasp of what these equations imply, let us study their asymptotic behaviors. 
Using  
\[ { N \choose p} \sim \frac{N^p}{p!},  N\rightarrow +\infty \]
and the assumption that $\Lfp{H^{p}}\ne 0$ lead to
\[ \Lfp{\E^N} \sim \frac{N^p}{p!} \Lfp{H^p}\]
and, with the assumption that $k$ diverges with $N$, to
\begin{equation}
\begin{split}
 \Prob_k(X_k= x) = \sum_{i,j\le p} {k-1 \choose i} {N-k \choose j} \frac{\Lfp{H^i R_d(x) H^j}}{\Lfp{\E^N}}  \\
  \sim \sum_{i+j=p} {p \choose i} \fracpow{k}{N}{i}  \left(1-\frac{k}{N} \right)^{j} \frac{\Lfp{H^j \Rd(x) H^{j}} } {\Lfp{H^{p}}}.
 \end{split}
\end{equation}  
Compared to \Eqref{eqn:margdiag}, this indicates that the marginal distribution of component $k$ reads as a mixture of distributions, with weights 
depending on the relative position $k/N$, rather than the absolute position $k$, as is the case for diagonalizable $\E$.

The covariance function also turns quite different from the diagonalizable case,
\begin{equation}
\begin{split}
 &\Esp {X_k X_l} \approx \sum_{i+j+m=p}\\
 &\left[ { \frac{p!}{i!m!j!} }\fracpow{k}{N}{i} \fracpow{k-l}{N}{j} \left(1-\frac{l}{N} \right)^{m} 
  \frac{\Lfp{H^i M_1 H^j M_1 H^{m}} } {\Lfp{H^{p}}} \right]
 \end{split}
\end{equation}  
with the occurrence of algebraic terms, $\fracpow{k-l}{N}{j}$, that indicate long-range correlations developing along the whole vector $\underline{X}_N$. 
 
This specific choice for non diagonalizable $\E$ enables us to figure out that, combining a block diagonal Jordan reduction of $\E$ and results obtained in the diagonalizable case, the covariance function consists, in the very general case, of a weighted sum of algebraic and exponential decreases. 

\section{Design and Synthesis}
\label{sec:synth}

The elegant and compact definition of $\X$ via its joint distribution $\Prob(\X)$ in Eq.~(\ref{eqn:prob}) gives little hints on how to construct random vectors with desired prescribed properties or on how to synthesize them numerically. 
The present section addresses such issues: First, given that $d, \A, \E$ and $\MP $ are chosen, Eq.~(\ref{eqn:prob}) is reformulated into the framework of Hidden Markov Models and the corresponding numerical synthesis algorithm is devised; Second, given $d, \A, \E$, an algorithm for constructing a Matrix $\MP$ with prescribed marginal distributions and dependence structures is detailed. 

\subsection{Hidden Markov Chain}
\label{sec:synth:hmc}

First, generalizing the fact that the entries of the matrix product $(ABC)$ reads $(ABC)_{i,j} = \sum_{k,l} a_{i,k}b_{k,l} c_{l,j}$, enables us to recast  Eq.~(\ref{eqn:prob}) into:
\begin{equation}
\label{eqn:synth_form}
 \Prob(x_1,\ldots,x_n)=  \sum_{\underline \Gamma }   \kappa(\underline{\Gamma}) \prod_{k=1}^{N} \GamF{\MP(x_k)}{k-1}{k} , 
\end{equation} 
with $ \underline{\Gamma} \equiv \{\Gamma_0, \ldots,\Gamma_k, \ldots, \Gamma_N \} \in [1,\ldots,d]^{N+1}$ and
\begin{equation} 
\label{eqn:kappa}
\kappa (\underline{\Gamma})= \frac{ \GamF{\A}{0}{N}}{\Lf (\E^N)} \prod_{k=1}^N \GamF{\E}{k-1}{k},
\end{equation} 
where $\sum_{\underline{\Gamma}} \kappa (\underline{\Gamma})=1$.
Therefore, $\kappa (\underline{\Gamma})$ and $\Prob(\Xv)$ can be read respectively as the probability function of $\underline{\Gamma}$ and
 as a $ \kappa (\underline{\Gamma})$-weighted mixture 
of laws, each defined as the product $\prod_{k=1}^N \MP_{\Gamma_{k-1},\Gamma_{k}}(x_k) $.

\begin{figure}
\begin{center}
   \def\svgwidth{6cm}
   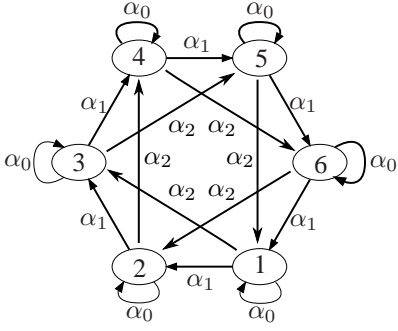
\end{center}
 
\caption{\label{Fig:markov} {\bf Transition graph.} Example for $d= 6 $ and $\E= \alpha_0 I_d + \alpha_1 J_d+\alpha_2 J_d^2$ ($\alpha_0+\alpha_1+\alpha_2 = 1$). 
} 
\end{figure}
\newcommand{\trGamma}[1]{\Gamma_{\ge #1}}

Second, let $\trGamma{t}$ denote the set of chains starting at index $t$ and stopping at index $N-1$: $\trGamma{t} \equiv (\Gamma_t,\dots,\Gamma_{N-1})$. For a given pair $(\Gamma_0,\Gamma_N)$, \Eqref{eqn:kappa} shows that: 
\begin{equation}
\begin{split}
 \Prob(\Gamma_{k}=j | \Gamma_{0}=\gamma_0,\ldots \Gamma_{k-1}=\gamma_{k-1}) = \\
  \frac{\E_{\gamma_0,\gamma_{1}} \ldots \E_{\gamma_{k-1}, j}  \sum_{\trGamma{k+1}}  \E_{j,\Gamma_{k+1}}  \ldots \E_{\Gamma_{N-1},\Gamma_N} } 
  {\E_{\gamma_0,\gamma_{1}} \ldots \E_{\gamma_{k-2},\gamma_{k-1}} \sum_{\trGamma{k}}  \E_{\gamma_{k-1}, \Gamma_{k}}  \ldots \E_{\Gamma_{N-1},\Gamma_{N} } } \\ 
 			              =   \E_{\gamma_{k-1},j}   \frac{\left(\E^{N-k}\right)_{j,\Gamma_N}}{\left(\E^{N-k+1}\right)_{\gamma_{k-1},\Gamma_N}}
 			               = \Prob(\Gamma_{k}=j | \Gamma_{k-1}=\gamma_{k-1})  
\end{split}
\end{equation}
and hence that $ \underline{\Gamma}$ consists of an inhomogeneous $d$-state Markov chain, with transition probability matrix at step $k$ reading: 
\begin{equation}
\label{eqn:transitions}
\Prob(\Gamma_{k}=j | \Gamma_{k-1}=i)= \E_{i,j} \frac{\left(\E^{N-k}\right)_{j,\Gamma_N}}{\left(\E^{N-k+1}\right)_{i,\Gamma_N}}.
\end{equation}
This shows that $\E$ can now be interpreted as (the basis for) the transition matrix underlying the Markov chain $\underline{\Gamma}$, 
as illustrated in Fig.~\ref{Fig:markov}.

Third, Eq.~(\ref{eqn:kappa}) enables us to show that the initial distribution for $(\Gamma_0,\Gamma_N)$ reads:
\begin{equation}
\label{eqn:initial_states}
\Prob(\Gamma_{0}=i, \Gamma_N=j)= \frac{\A_{i,j} (\E^N)_{i,j}} {\Lfp{\E^N}}.
\end{equation}

Combined together, these three steps enable us to synthesize numerically Vector $\Xv_k$, as defined from Eq.~(\ref{equ-Rd}), using a Hidden Markov Model, with $4$ hidden states: 
the current state $\Gamma_k$, the previous state $\Gamma_{k-1}$ and the pair formed by the initial and final states $(\Gamma_0,\Gamma_N)$. 
Combining Eqs.~(\ref{eqn:transitions}) and (\ref{eqn:initial_states}), the synthesis algorithm can be sketched as follows: 
\begin{enumerate}
\item[Step1:] Initialization: \\
Use Eq.~(\ref{eqn:initial_states}) to generate the states $\Gamma_0$ and $\Gamma_N$. 
\item[Step2:] Iteration on $k=1, \ldots, N$: 
\begin{enumerate}
\item[Step2.1:] Choose at random state $\Gamma_k$, according to the transition probability given in Eq.~(\ref{eqn:transitions}); 
\item[Step2.2:] Generate $X_k$ according to $\MP_{\Gamma_{k-1},\Gamma_{k}}$.
\end{enumerate}
\end{enumerate}
\vspace{2mm}

Under the particular choices $\A_{i,j} = 1/d$  (cf. \Eqref{equ:Astat}) and $ \E $ doubly stochastic  (cf. \Eqref{equ:EDS}),
it has been shown that there is no need to set a priori the final state $\Gamma_N$ and that the Markov chain becomes homogeneous, hence that Eqs. (\ref{eqn:transitions}) and (\ref{eqn:initial_states}) can be rewritten as (cf. \cite{Angeletti2012:ICASSP}):
\begin{equation}
\label{eqn:transitions:sto}
\Prob(\Gamma_{k}=j | \Gamma_{k-1}=i)= \E_{i,j} \makebox{ and }
\end{equation}
\begin{equation}
\label{eqn:initial_states:sto}
\Prob(\Gamma_{0}=i)= \frac{1} {d}.
\end{equation}
Such simplifications show that $\E$ exactly defines the transition matrix for the Markov chain $\Gamma$ and, 
moreover, that the initial state of the Markov chain $\Gamma_0$ follows a uniform distribution. 

The computational cost of the synthesis procedure can be evaluated as follows. 
For the general case where $\E$ is not an invertible matrix, the algorithmic complexity is dominated by the cost of the computation of matrix powers, and hence a total complexity in $\mathcal{O}(d^3 N \ln N )$ (with a $\mathcal{O}(d^3)$  matrix multiplication cost and a $\mathcal{O}(\ln N)$ matrix power computation cost). 
When $\E$ is invertible, the global cost reduces to $\mathcal{O}(d^3 N)$ (as matrix multiplications only are required). 
With the choice of doubly stochastic matrices $\E$ and $\A_{i,j}=1/d$, the computational cost is further reduced to $\mathcal{O}(N \ln d)$. 
Therefore, the synthesis procedure scales efficiently for large signal (large $N$) and large complex dependence structure (large $d$) (with numerous further potential optimizations for specific cases such as sparse structure matrix).

\subsection{Design}
\label{sec:invprob}

Let us now address the issue of how to select the elements constitutive of the model, given a targeted $\X$. 

Prescribing directly a targeted joint distribution function consists of a difficult theoretical problem, that also often does not match application purposes. 
Instead, it is often natural to split this general question into two simpler sub-problems: on one hand, fixing the dependence (or covariance) structure; on the other hand, fixing the marginal distribution(s).
Sections~\ref{sec:theory} and \ref{sec:asympt} indicate that ${ r_M \choose 2}$ defines the (maximal) number of time-scales involved in the dependence (covariance) structure function, while the joint choice of $\A$ and $\E$ controls the shape of the dependence (e.g., stationarity, cross-covariance, \ldots). 
Examples of construction enabling us to reach targeted dependencies are detailed in Section~\ref{sec:illust}. 

Let us for now assume that $d, \A$, and $ \E$ are chosen (hence that the dependence structure is fixed) and let us concentrate here on designing Matrix $\MP$, so as to reach targeted marginal distributions. 
Also, for the sake of simplicity, let us concentrate on the stationary case first. 
Extensions to the non-stationary case will then be discussed. 

It is first worth emphasizing that prescribing marginals and dependence cannot be treated as fully independent problems (only the global structure of the dependence is not related to the choice of Matrix $\MP$).
Indeed choosing $\MP$ fixes both univariate distributions (according to \Eqref{eq:inv:univ}) and the coefficients governing the relative amplitude of the dependence structure, through the matrices $\Mq{q}$ (cf.  \Eqref{eqn:moments:multi}). 
Moreover, the prescribed marginal imposes some constraints on the matrix $\Mq{q}$.
For instance, \Eqref{eqn:static:updf} implies that:
\begin{equation}
\Mq{1}_{i,j}  <  \E_{i,j} \Esp{X| X > x}, \quad x\equiv F_S^{-1}(1-c_{i,j}),
\end{equation}
with $X$ generated according to $\Prob_S$ (as defined in \Eqref{eqn:static:updf}), and where $F_S$ denotes the cumulative function associated to the marginal distribution $\Prob_S$.
The difficulty thus consists of devising a constructive procedure for Matrix $\MP$ enabling practitioners to reach jointly the targeted marginal distribution $\Prob_S$ (according to \Eqref{eqn:static:updf})  and  targeted Matrices $\Mq{q}$ (cf.  \Eqref{eqn:moments:multi}). 
To disentangle both groups of constraints, the $\MP_{i,j}$ are parametrized using a matrix of (strictly) positive functions $g_{i,j}(x)$:
\begin{equation}
\label{equ:amu}
c_{i,j} \MP_{i,j}(x) = \frac{ g_{i,j}(x)}{\sum_{l,m} g_{l,m}(x) } \Prob_S(x).
\end{equation}
With this parametrization, \Eqref{eqn:static:updf} is automatically satisfied.
To ensure that the $\MP_{i,j}$ are probability distributions, it is needed that
\begin{equation}
\label{equ:amub0}
\int  \frac{g_{i,j}(x)}{\sum_{l,m} g_{l,m}(x) } \Prob_S(x) dx = c_{i,j}.
\end{equation}
Splitting $g_{i,j}$ into $g_{i,j}(x)=\mu_{i,j} h_{i,j}(x)$ permits to use the free parameters $\mu_{i,j}$ to solve \Eqref{equ:amub0}, for any fixed $h = (h_{i,j})$ :
\begin{equation}
\label{equ:amub}
\int  \frac{\mu[h]_{i,j} h_{i,j}(x)}{\sum_{l,m} \mu[h]_{l,m} h_{l,m}(x) } \Prob_S(x) dx = c_{i,j}.
\end{equation}
Plugging the solution $(\mu[h]_{i,j})$ of \Eqref{equ:amub} into  \Eqref{equ:amu} yields:
\begin{equation}
\label{equ:amu_h}
\MP[h]_{i,j}(x) = \frac{\mu[h]_{i,j} h_{i,j}(x)}{c_{i,j}\sum_{l,m} \mu[h]_{l,m} h_{l,m}(x) } \Prob_S(x).
\end{equation}
The last step is to find $h$ such that:
\begin{equation}
\label{equ:amuc}
\Mq{q}_{i,j} = \E_{i,j} \int x^q \MP[h]_{i,j}(x)  dx
\end{equation}
where $\Mq{q}$ are the prescribed moment matrices.
Essentially, this parametrization of $ \MP  $ in terms of $h_{i,j}(x)$  enables us to recast \Eqref{eqn:Mq} as a non linear equation in $h_{i,j}(x)$ (cf. \Eqref{equ:amuc}), where $h_{i,j}(x)$ are strictly positive functions. 
To practically find solutions to Eq.~(\ref{equ:amuc}), the functions $h_{i,j}(x)$ are further constrained to be chosen amongst an arbitrary parametrized kernel family $K_p$: $h_{i,j}(x) = K_{p_{i,j}}(x)$. A natural choice for the kernel parameters $p$ would be to consider vectorial parameters with a dimension equal to the number of targeted moments orders.
For instance, when the targeted $\Prob_S$ is a normal law and $\Mq{1}$ and $\Mq{2}$ are fixed a priori, it is natural (but not mandatory) to select $K_{m,\sigma} = \Normal{m}{\sigma}$.
Further examples are given in Section~\ref{sec:illust}. 
Within this framework, Eqs.~(\ref{equ:amu}) to (\ref{equ:amuc}) can be solved numerically step-by-step.
Examples of results obtained with the constructive method are detailed in Section~\ref{sec:illust} and illustrated in Fig.~\ref{fig:trivariate:univ}.
 
In summary, to design Vector $\X$ with prescribed univariate distributions and dependence structure, the following procedure is to be followed: 
\begin{enumerate}
\item{Select $d$, $\E$ and $\A$ in agreement with the targeted dependence structure.}
\item{Choose the moment matrices $\Mq{q}$ to determine the dependence coefficients.}
\item{Choose a kernel family $K_p$.}
\item{Solve Eqs.~(\ref{equ:amu}) to (\ref{equ:amuc}) to compute the distribution matrix $\MP$ in agreement with the targeted $\Prob_k$s.}
\end{enumerate}
A {\sc Matlab} implementation of this procedure is available upon request.
The potential of this method is further explored in Section \ref{sec:illust} where simple (hence pedagogical) examples are devised and discussed.

For the general non-stationary case, the above algorithm cannot be used as is and a general solution is still under construction.
However, for the restriction of the non-stationary case to the fairly broad and practically efficient case of block structured matrices, detailed in Section~\ref{sec:mvr}, 
the above algorithm can straightforwardly be applied independently for each component of the targeted random vector. 
As illustrated in Section~\ref{sec:mvr}, both the block structured matrix framework and the above algorithm provide practitioners with a rather general tool to define and synthesize numerically a large number of realizations of random vectors.

\section{Illustrations}
\label{sec:illust}

This sections aims at illustrating the potential of the general construction above in two specific contexts: First, we concentrate on stationary vectors $\X$, i.e., on stationary time series synthesis; Second,  
we illustrate for trivariate random vectors how to vary the joint distribution functions while maintaining fixed the three different marginal distributions. 

\subsection{Stationary time series}
\label{sec:static}

\subsubsection{Circular structure matrix}
\label{sec:circular}

To design stationary time series, elaborating on a preliminary work (cf. \cite{Angeletti2012:ICASSP}),
we now select $\A_{i,j}= (1/d)$ and a particularly fruitful choice of doubly stochastic matrices: 
\begin{equation}
\E =  \sum_{k=0}^{d-1} \alpha_k J^k _d, \, \, \, \sum_{k=0}^{d-1} \alpha_k \equiv 1, 
\end{equation}
where $J_d \in M_d(\R) $ is defined as: 
\[ J_d = 
\begin{pmatrix}
0       & 1      &0       & \cdots  & 0      \\
\vdots  & \ddots & \ddots & \ddots  & \vdots \\
\vdots  &        & \ddots & \ddots  & 0      \\
0       &        &        & \ddots  & 1      \\
1       & 0      & \cdots & \cdots  & 0 \\  
\end{pmatrix}
\]
and $J_d^0 = I_d$, the Identity matrix.
It is easy to verify that $[\A^T,J_d]=0$. 
Therefore, $[\A^T,\E]=0$ and $\X$ is stationary. 
Moreover, the eigenvalues of  $J_d$ are easily computed as functions of the roots of order $d$ of the unity, $\omega= \exp( 2 \imath \pi / d)$, which motivates the use of this matrix: 
\begin{equation}
\lambda_i =  \sum_{k=0}^{d-1} \alpha_k \omega^{ik}, \, i=0, \ldots, d-1.
\end{equation}
The eigenvectors of $J_d$ also have simple expressions:
\begin{equation}
B_{i,j}= \omega^{i j}.
\end{equation}
From these definitions, it is useful to change of basis: 
\begin{equation}
\begin{gathered}
\At= B \A B^{-1} = \begin{pmatrix}
0      & \hdots  & 0      \\
\vdots &         & \hdots \\
0      & \vdots  & 1      \\
\end{pmatrix} \\
\tilde{\E}= B^{-1} \E B=
\begin{pmatrix} 
\lambda_1& &0\\
&  \ddots &\\
0& & \lambda_d \\ 
\end{pmatrix} \\
\Mqt(q)=  B^{-1} \Mq{q} B
\end{gathered}
\end{equation}
Such calculations lead to a simple expression of the dependencies:
\begin{equation}
\Esp{X_k^{q_1} X_l^{q_2}} = \sum_{i} \lambda_i^{k-l} \Mqt(q_1)_{d,i} \Mqt(q_2)_{i,d}
\end{equation}
For ease of notations, let us define the two vectors $ (C_{\Mqt(q)})_k = \sum_i \Mqt(q)_{i,k}$,  $(R_{\Mqt(q) })_k= \sum_j \Mqt(q)_{k,j} $ and let $\F$ denote the (normalized) discrete Fourier transform :
\begin{equation*}
\label{eqn:def:DFT}
\F( v )_k = \frac{1}{d}\sum_{l=1}^{d} v_l e^{\frac{2\imath \pi k l}{d}}= \sum_{l} v_l \omega^{kl}
\end{equation*}
This leads to: 
\begin{equation}
\label{eqn:eigen:fft}
\begin{split}
&\Esp {X_0 ^q X_t^q} -\Esp{X_0^q} \Esp{X_t^q}=\\
& \sum_{k=1}^{\lfloor d/2 \rfloor }  m_k \Real  { \F(R_{\Mqt(q)})_k  \overline{\F(C_{\Mqt(q)})_k } e^{-\frac{t-1}{\tau_k}} e^{ \imath \frac{2 \pi (t-1)}{T_k} }}, 
\end{split}
 \end{equation}
where $\lfloor z \rfloor $ stands for the integer part of $z$, $m_k = 1 $ if $ 2k=d$ and $m_k=2$ otherwise, and $\Real{} $ denotes the real part. In this specific case, there is thus at most $ \lfloor d/2 \rfloor +1$ distinct time scales in the signal.
As a pedagogical example, choosing $\E = \alpha_0 I+\alpha_1 J_d$ (where $\alpha_0+\alpha_1=1$) enables to further study the form of the covariance function. 
The eigenvalues of $\E$ read $ \lambda_k= \alpha_0 + \alpha_1 e^{\frac{2 \imath \pi k}{d}} $ and can be rewritten as $ \lambda_k= e^{-\frac{1}{\tau_k}} e^{\pm \imath \frac{2 \pi}{T_k} } $,  $k=1, \ldots \lfloor d/2 \rfloor$. 
Combining these results with \Eqref{eqn:eigen:fft} shows that $\tau_k = - 1/ \ln  |\lambda_k| $ and $T_k = 2 \pi / \arg(\lambda_k) $ are characteristic dependence time scales and periods that depend on the joint choice of $d$ and $ \alpha_0 $. 
It can also be shown that $\tau_k \approx_{\alpha_0 \alpha_1 \rightarrow 0} \left[ \alpha_0 \alpha_1 \left(1- \cos \frac{2 \pi k}{d} \right) \right]^{-1}$.
Moreover, increasing $d$ increases both the number of distinct dependence time scales and the ratio of the smallest to the largest such characteristic time scales, which can be shown to vary asymptotically ($\alpha_0 \alpha_1 \rightarrow 0$) as $(d/2\pi)^2/2$.

\newcommand{\sigA}{\sigma_1}
\newcommand{\sigB}{\sigma_2}
\newcommand{\Yv}{\underline{Y}}
\subsubsection{Example}
\begin{figure}[htb]
\centerline{\includegraphics[width=0.49\linewidth,height=25mm]{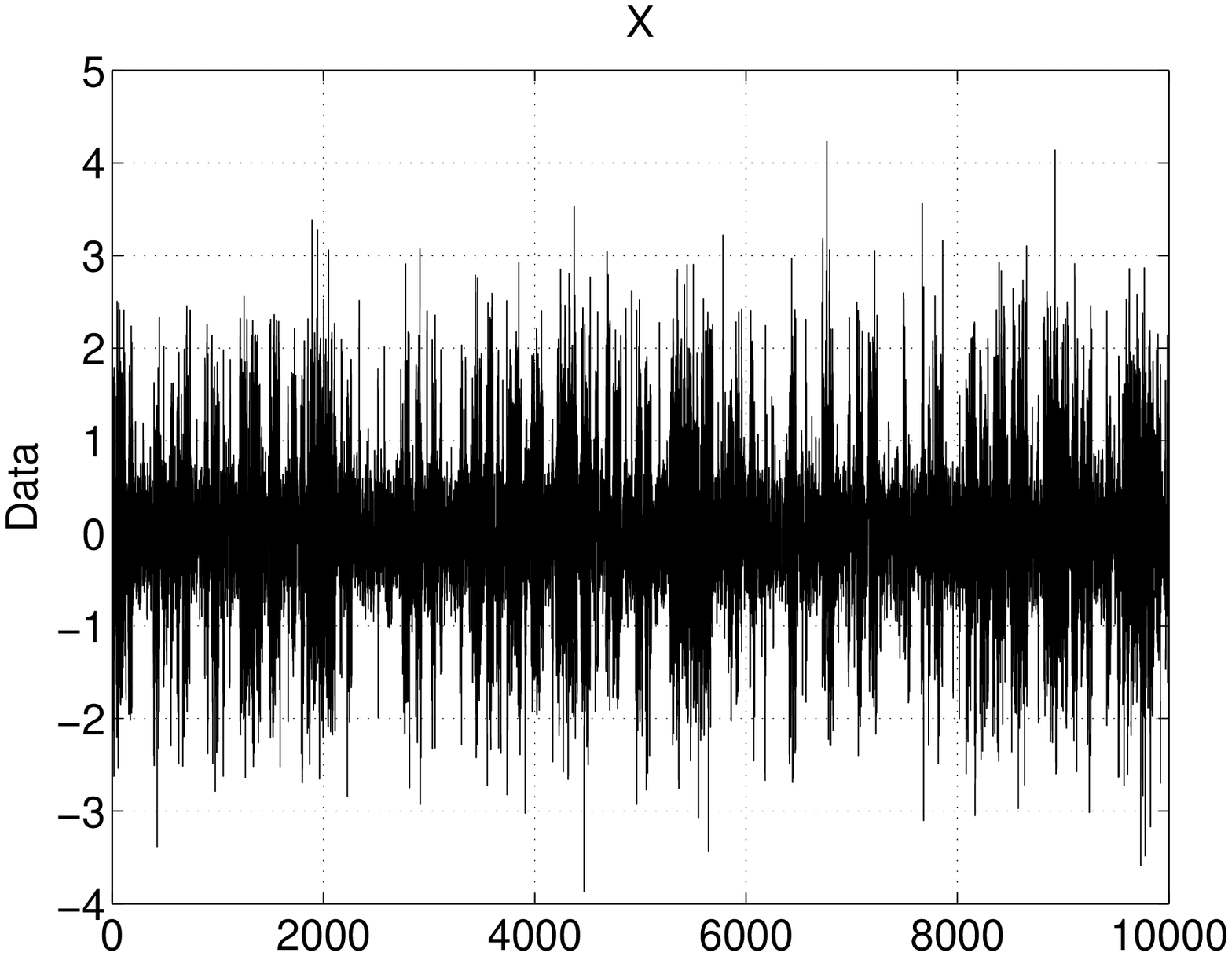} 
\includegraphics[width=0.49\linewidth,height=25mm]{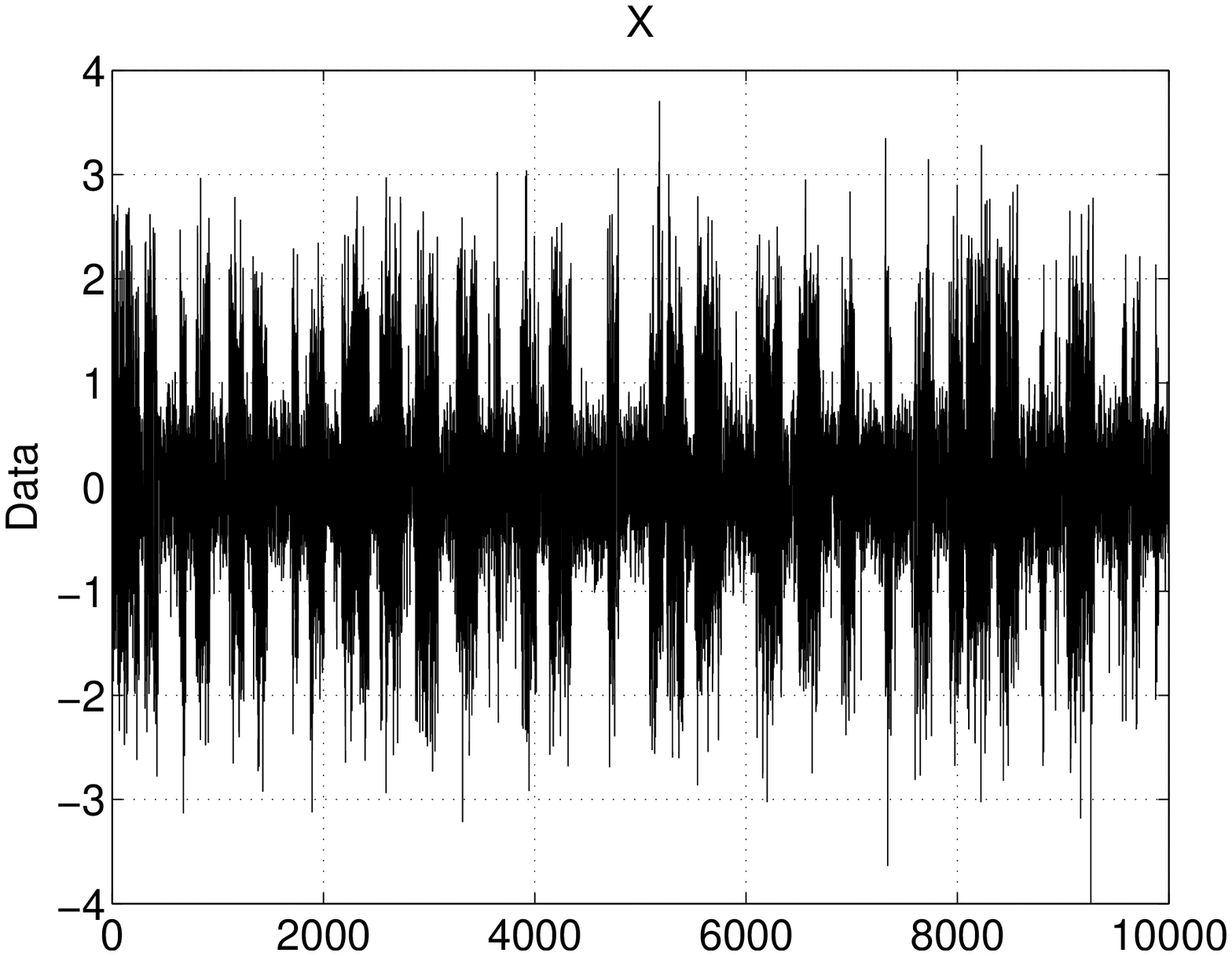}}
\centerline{\includegraphics[width=0.49\linewidth,height=25mm]{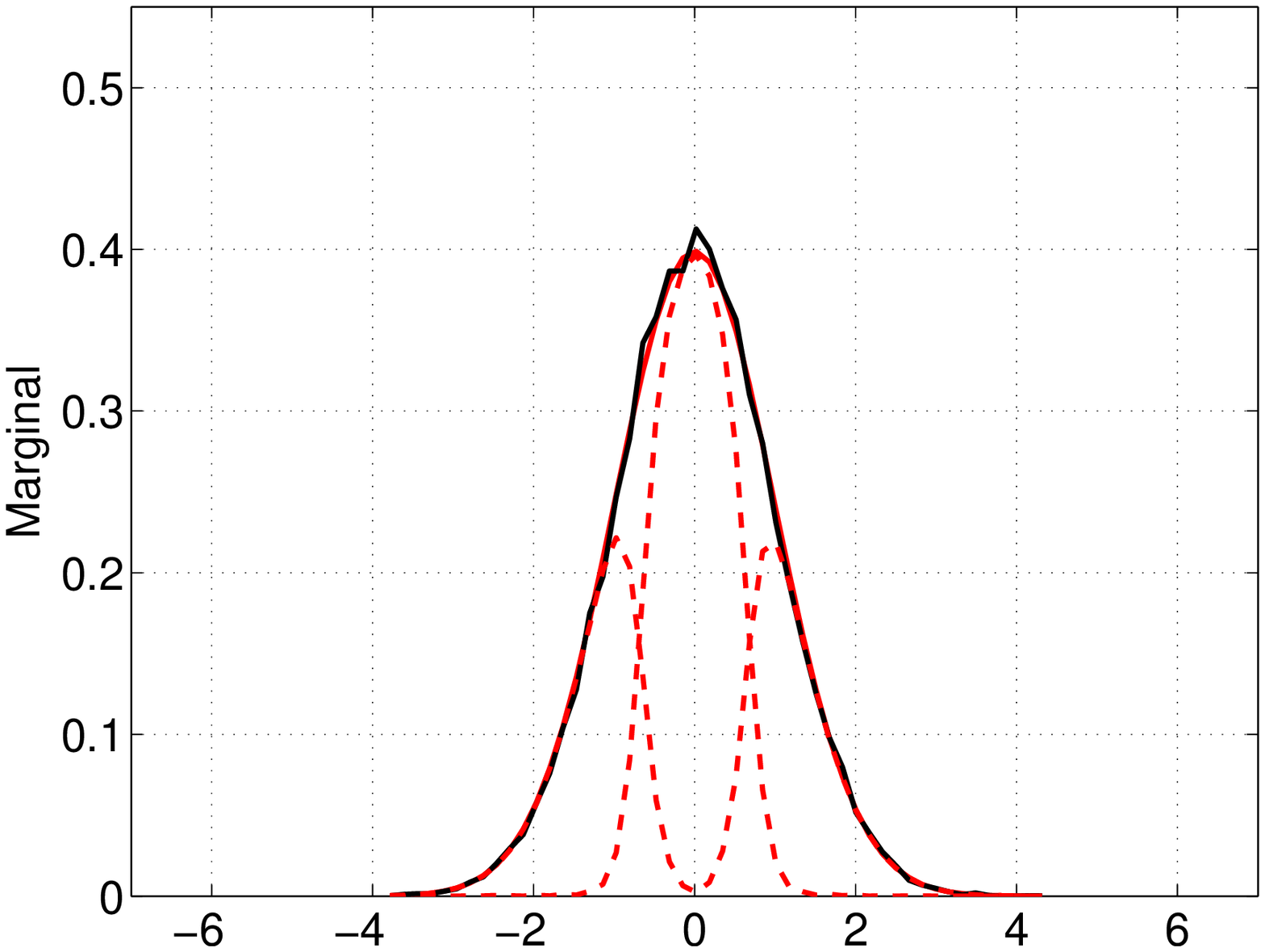}
\includegraphics[width=0.49\linewidth,height=25mm]{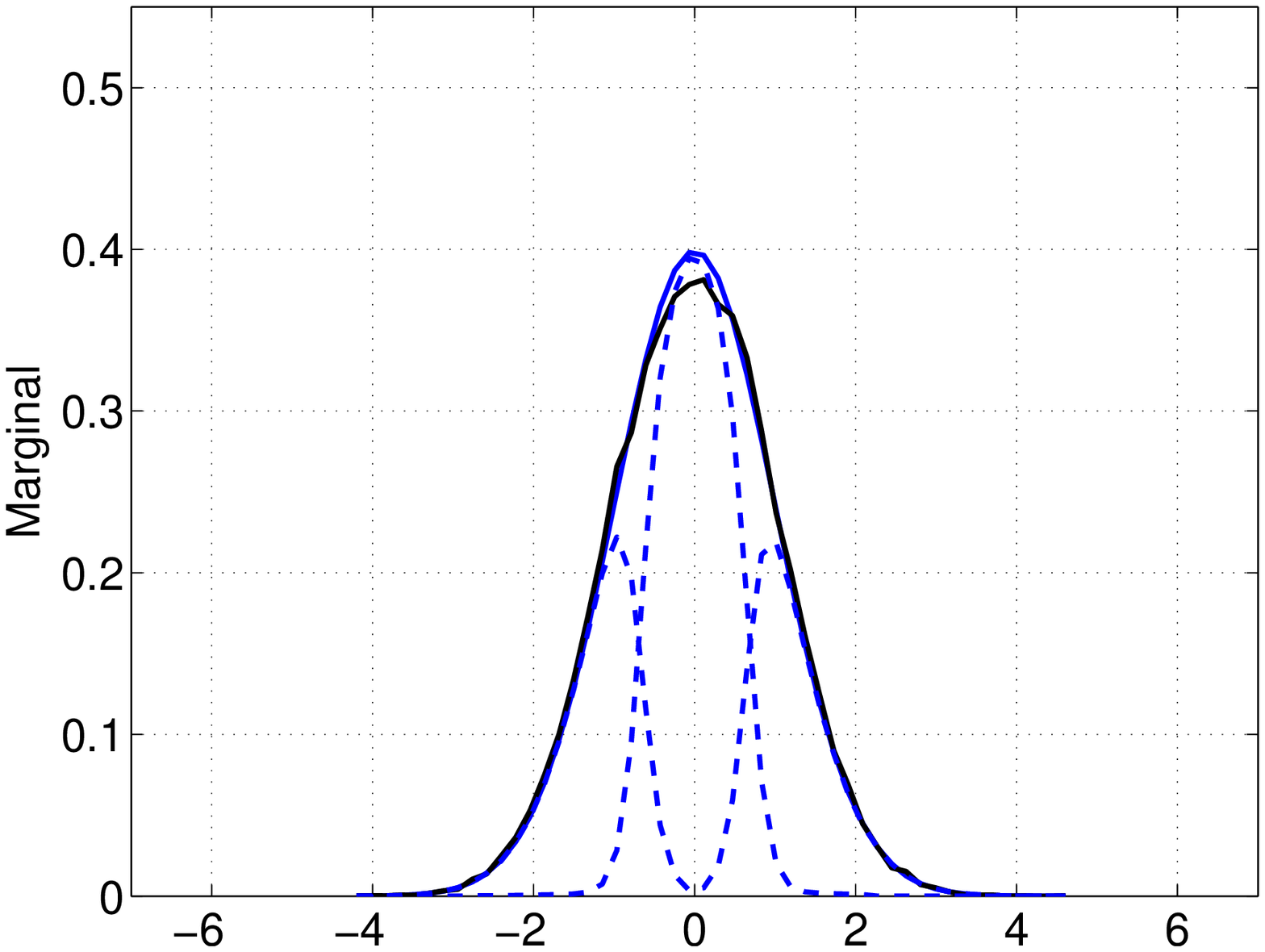}}
\centerline{\includegraphics[width=0.49\linewidth,height=25mm]{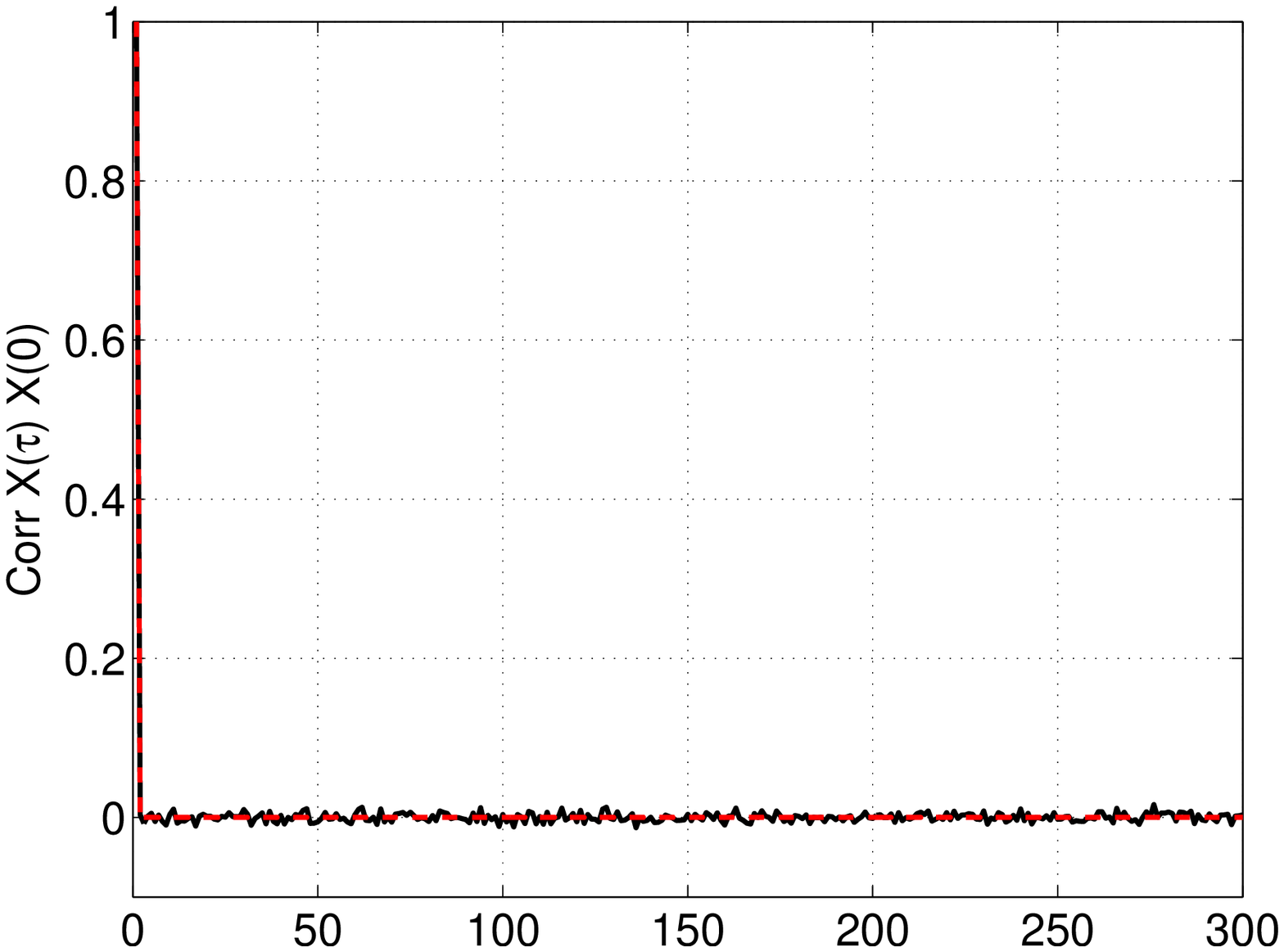}
\includegraphics[width=0.49\linewidth,height=25mm]{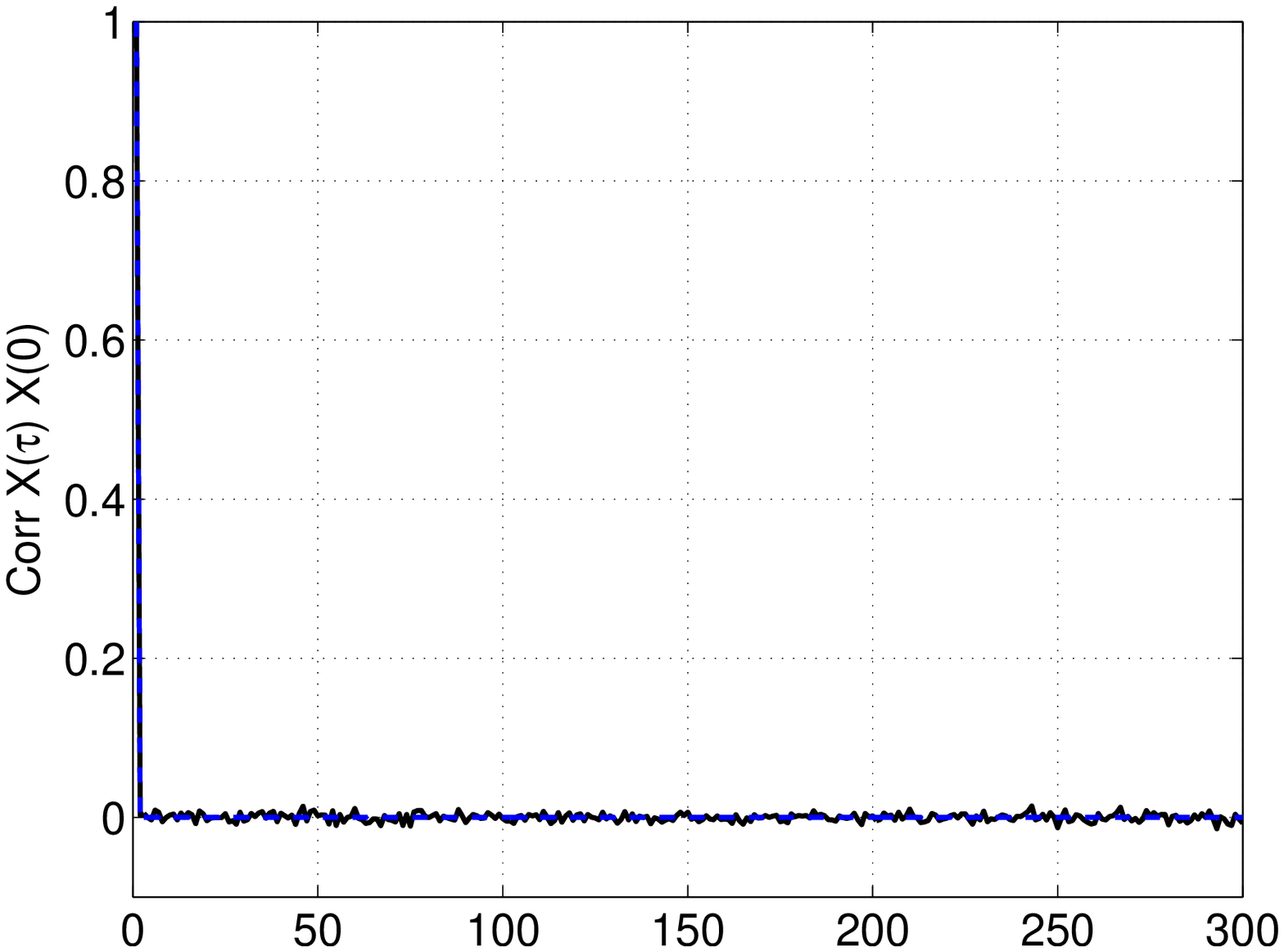}}
\centerline{\includegraphics[width=0.49\linewidth,height=25mm]{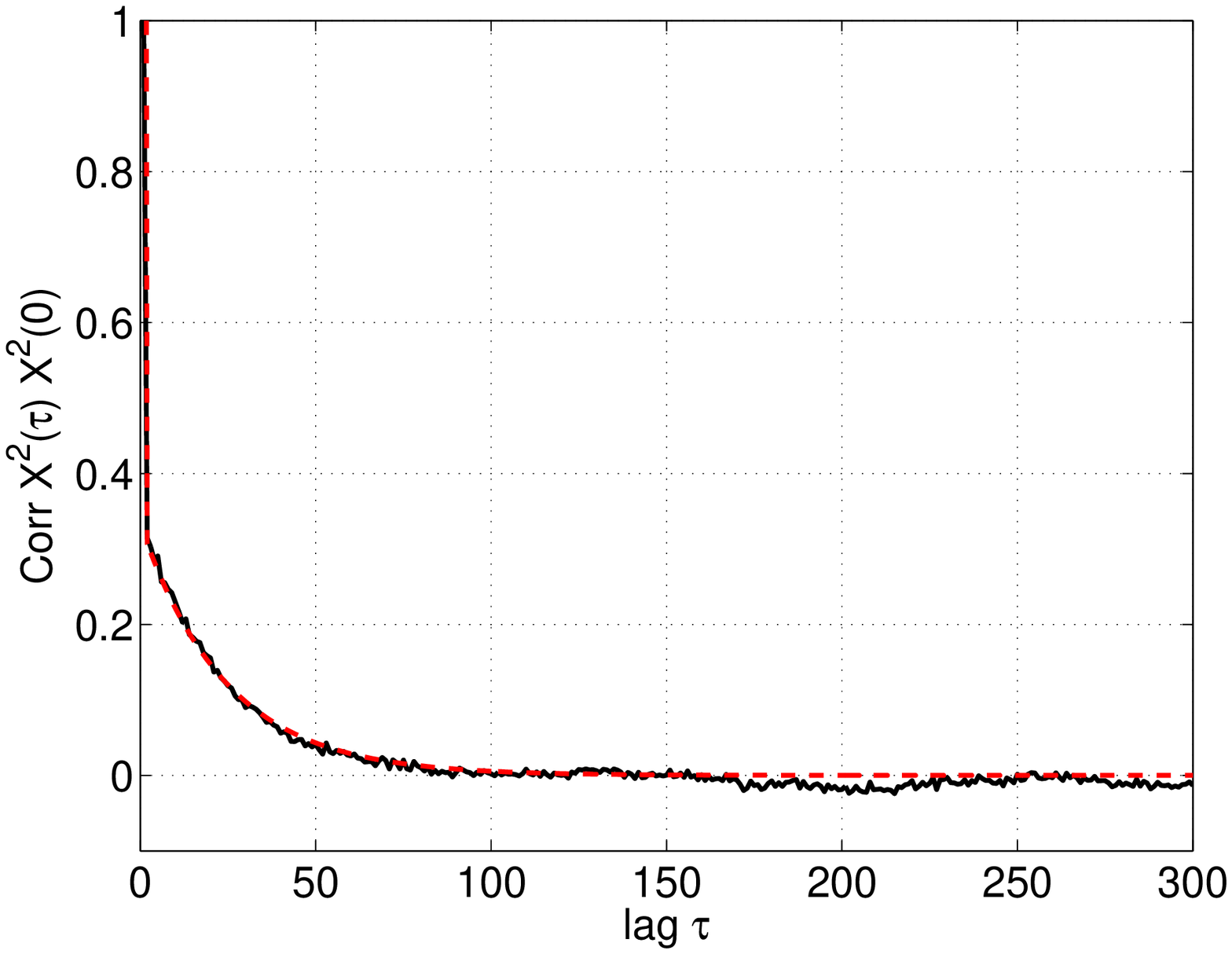}
\includegraphics[width=0.49\linewidth,height=25mm]{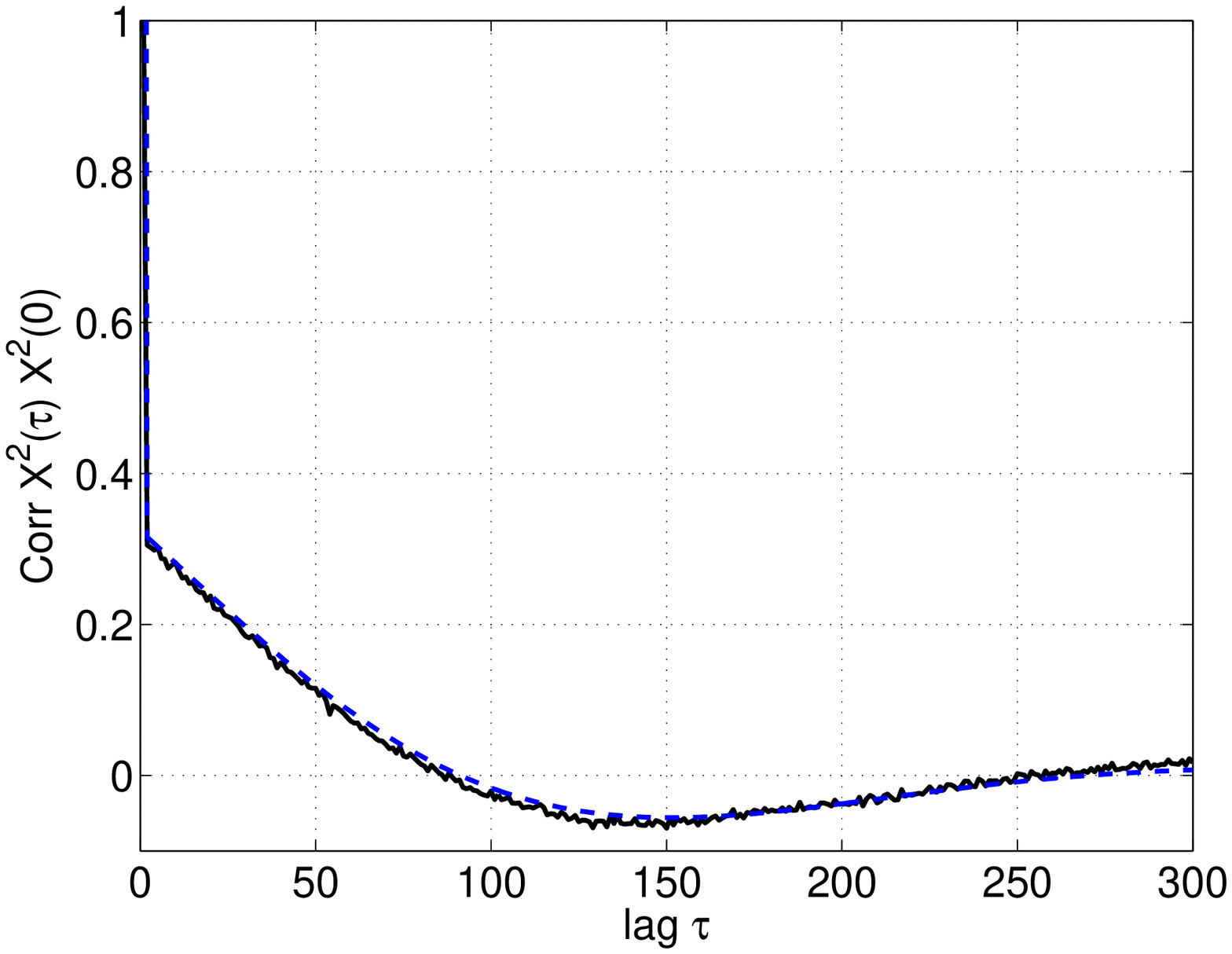}}

\caption{\label{fig:results} {\bf Numerical Synthesis.} Two different time series with the same marginal distribution (mixture of two Gaussians, $\sigA=0.5$, $\sigB=2$), the same covariance functions (chosen as $\delta$ functions) but different covariance functions for their squares, hence different prescribed joint distributions ($\alpha_0=0.98$). Left side $\Xv$, right side $\Yv$. First line, one realization of the time series. Then, from top to bottom, estimated (solid black lines) and theoretical (dashed colored lines) marginals, correlation functions and correlation functions for the squared time series.}
\end{figure}

To illustrate the potential of the proposed time series theoretical construction and synthesis procedure, a pedagogical example is devised.
First, it is intended to design a stationary Gaussian time series $X$ of size $N$, with a Gaussian marginal distribution $P(x) = \Normal{0}{1}$, an autocovariance function that consists of a $\delta$ function (i.e., no correlation and $\Mq{1} \equiv 0$), but dependence as fixed here by the $4-$th order statistics, i.e., by prescribing $\Mq{2} $.
By definition, $X$ is hence not a jointly Gaussian process (i.e., its joint distribution departs from a multivariate Gaussian law). 
Second, it is intended, using the same $d$, $\A$ and $\E$, to design a second time series $Y$, independent of $X$, but with same size $N$, same marginal distribution and autocovariance function but different dependence (via a different $\Mq{2}$ matrix) and hence a joint distribution different from that of $X$.

For these constructions, $d=6$ is selected to obtain $3$ distinct time scales and $\E= \alpha_0 I_d + \alpha_1 J_d$ ($\alpha_0+\alpha_1=1$) for the sake of simplicity. 
With these choices, the three available time scales read:

\begin{equation}
 \tau_1 \underset{\alpha_1\rightarrow 0}{\approx} \frac{2}{\alpha_1},\, 
 \tau_2 \underset{\alpha_1\rightarrow 0}{\approx} \frac{2}{3 \alpha_1},\, 
 \tau_3 \underset{\alpha_1\rightarrow 0}{\approx} \frac{1}{2 \alpha_1}.
 \end{equation}

The two time series $X$ and $Y$ differ only in the autocovariance of their squares ($\Esp{\Xv^2_0 \Xv^2_t} - \Esp{\Xv^2_0}\Esp{\Xv^2_t}$). 
This implies that the matrices $\Mq{2}$ of the two signals differ. 
An interesting simplification at this step is to choose the matrices $\Mq{2}$ to be of the form $\Mq{2}=, \alpha_0 D + \alpha_1 D J_d$ where $D$ is diagonal. 
In this case \Eqref{eqn:eigen:fft} further simplifies to:  
\begin{equation}
\label{eqn:eigen:fft:diag}
\begin{split}
&\Esp {X_0 ^q X_t^q} -\Esp{X_0^q} \Esp{X_t^q}=\\
& \sum_{k=1}^{\lfloor d/2 \rfloor }  m_k |\F(\fun{diag}(D))_k|^2 \Real  { (\alpha_0 +\alpha_1 \omega^k) e^{-\frac{t-1}{\tau_k}} e^{ 2 \pi \imath \frac{t-1}{T_k} }}, 
\end{split}
 \end{equation}  
 where $diag(D)_i = D_{i,i}$ is the diagonal vector of $D$.
Therefore, choosing for $X$, 
\begin{equation}
 D_{i,i}=
\begin{cases} \sigma_1 & \text{i even} \\
              \sigma_2 & \text{otherwise} \\                             
\end{cases}, \quad \sigma_1+\sigma_2=1
\end{equation} 
induces that $\tau_3$ only, i.e., the shortest time scale, appears in the auto-covariance of $X^2$: \begin{equation}
 \Esp{X_0^2 X_t^2}-\Esp{X_0^2}\Esp {X_t^2}= \frac{(\sigA -\sigB)^2}{4}(\alpha_0-\alpha_1)^t.
 \end{equation} 
Conversely, imposing for $Y$
\begin{equation}
D_{i,i} = 
\begin{cases} \sigma_1 & i \le 3 \\
              \sigma_2 & i > 3 \\                             
\end{cases}, \quad \sigma_1+\sigma_2=1
\end{equation} 
leads to the fact that both $\tau_1$ and $\tau_3$ contribute to the autocovariance of $Y^2$:
\begin{equation}
\begin{aligned}
&\Esp{Y_0^2 Y_t^2}-\Esp{Y_0^2}\Esp {Y_t^2}=\\ \frac{(\sigA -\sigB)^2}{36} &\left[(\alpha_0-\alpha_1)^{t}+ 4 \Real{(2-\alpha_1-\sqrt{3} \imath \alpha_1) \lambda_1^{t-1} } \right].
\end{aligned}
\end{equation}
However, $\tau_1 \approx 4 \tau_3$ is dominant at large $t$ and hence constitutes the leading term. 

With these pedagogical and documented choices of $\Mq{2}$ for $X$  and $Y$, we can use the numerical procedure devised in Section~\ref{sec:invprob}, with a Gaussian kernel $K_{m,\sigma}(x)= \exp(-(x-m)^2/(2\sigma^2))$, to compute the associated distributions $\MP_{i,j}$. The hidden Markov chain procedure described in Section~\ref{sec:synth:hmc} is then used to synthesize the signals $X$ and $Y$. The analysis of these synthesized signals produces the results reported in Fig.~$\ref{fig:results}$.
This figure presents for $X$ (left column) and $Y$ (right column) a particular realization of the time series, the estimated and targeted univariate distributions, covariance functions and covariance functions for the squared time series (from top to bottom). 
It clearly shows that $ X $ and $ Y $ have the same marginal and covariance but different joint distributions (as targeted).
Though sharing the same pedagogical goal, the examples devised here significantly differ from those presented in \cite{Angeletti2012:ICASSP} as both the targeted marginal and the constructive Kernel differ. 

Using the same construction procedure, other examples with the same (non necessarily Gaussian) marginals, same (non necessarily $\delta$-) autocovariance functions but different joint distributions could as easily be devised and are available upon request.
 
\subsection{Random vector}
\label{sec:mvr}

\subsubsection{Multivariate design}

Let us now consider the design of a random vector whose components have different univariate distributions and are dependent. 
As seen in Section \ref{sec:asympt}, for a fixed $d$, when $N$ increases, $\X$ tends to become stationary. 
A simple way around this drawback is to increase the size of $\Rd$ with the size of $\X$, by choosing $d=(N+1) d^*$, while  
keeping a block triangular superior structure, to avoid a too large increase in the number of degrees of freedom:
\newcommand{\Rds}[1]{R_{d^*}^{(#1)}}
\newcommand{\Estl}{{\E^*}}
\newcommand{\Es}[1]{\E^{(*#1)}}

\newcommand{\MPs}[1]{\MP^{(*#1)}}
\begin{equation}
\label{equ:rds}
R_{d}(x) = 
\begin{pmatrix}
0_{d^*}& \Rds{1}(x)& & 0_{d^*} \\
& 0_{d^*} & \ddots &\\
& & 0_{d^*} &  \Rds{N}(x) \\
0_{d^*} & & & 0_{d^*}
\end{pmatrix},
\end{equation}
where $0_{d^*}$ is a $0$-block of size $d^* \times d^*$ and $\Rds{k}$ are $d^*\times d^*$ matrices, as defined in \Eqref{equ-Rd}: 
\[ \Rds{k}(x) = \Es{k} \MPs{k}(x), \quad \int_{-\infty}^{+\infty} \MPs{k}_{i,j}(x) dx=1. \]
Let us also choose the projection matrice $\A$ such that 
\begin{equation}
\label{equ:astar}
\A =
\begin{pmatrix}
0_{d^*} &\dots & 0_{d^*}\\
\vdots & & \vdots\\
\A^*& \dots& 0_{d^*} \\
\end{pmatrix},
\end{equation}
where $\A^*$ is a positive entry matrix of size $d^* \times d^*$. 
If one further defines $\Lfsp{M}=\Tr{\A^{*T} M}$ (as a linear form on $d^* \times d^*$ matrices),  \Eqref{eqn:prob} becomes:
\begin{equation*}
\Prob(x_1,\dots,x_N)=\frac{\Lfsp { \prod_{k=1}^N \Rds{k}(x_k) } }{\Lfsp{ \prod_{k=1}^N \Es{k} } }. 
\end{equation*} 
By further restricting to the case $\Es{k}=\Estl$, the joint distribution simplifies to: 
\begin{equation}
\label{eq:prob:nstatic}
\Prob(x_1,\dots,x_N)=\frac{\Lfsp { \prod_{k=1}^N \Rds{k}(x_k) } }{\Lfsp{\Estl^N } }. 
\end{equation}

The joint distribution in \Eqref{eq:prob:nstatic} consists of a variation on \Eqref{eqn:prob}, where the constant probability matrix $\MP$ has been replaced by a varying probability matrix $\MPs{k}$. 
The general formulation in \Eqref{equ:rds} and \Eqref{equ:astar} shows that this particular setting is nothing but a convenient notation that however corresponds to a subcase of the general framework: Therefore all results developed in Sections~\ref{sec:theory} to \ref{sec:synth} remain valid.

It is hence straightforward to derive the corresponding expressions for the univariate distributions, the $p$-samples probability distributions and moments:

\begin{equation}
\label{eqn:pdf:nstatic:univ}
\begin{split}
\Prob_k(X_k=x_k)= \frac{1}{\Lfp{\Estl^N}}
\,  \Lfp{\Estl^{k_1-1} \Rds{k}(x_{k}) \Estl^{N-k} }
 \end{split}
\end{equation}

\begin{equation}
\label{eqn:pdf:nstatic}
\begin{split}
&\Prob(X_{k_1}=x_{k_1}, \dots, X_{k_p}=x_{k_p})= \frac{1}{\Lfp{\Estl^N}}\\
&  \Lfp{\Estl^{k_1-1} \left( \prod_{r=1}^{p-1} \Rds{k_r}(x_{k_r}) \Estl^{k_{r+1}-k_{r}-1}  \right) \Rds{k_p}(x_{k_p}) \Estl^{N-k_{p}} }
 \end{split}
\end{equation}
\newcommand{\Mqs}[2]{M^{(*#1)}\left(#2 \right)}
\begin{equation} 
\label{eqn:moments:nstatic}
\begin{split}
&\Esp{\prod_{r=1}^p X_{k_r}^{q_r}}= \frac{1}{\Lf(\Estl^N)}\\
& \Lf \left(\Estl^{k_1-1} \left( \prod_{r=1}^{p-1} \Mqs{k_r}{q_r} \Estl^{k_{r+1}-k_{r}-1}  \right)
 \Mqs{k_p}{q_p} \Estl^{N-k_{p}} \right) ,
 \end{split}
\end{equation}
with $\Mqs{k}{q}= \int_{-\infty}^{+\infty} x^q \Rds{k}(x) dx$.
Using this particular structure for the matrix $\Rd$ permits to define each component of the random vector $\X$ with a relative independence. Notably, the univariate distribution of $X_k$ depends only on $\MPs{k}$.
Moreover, with these matrices $\A$ and $R_d$, the hidden Markov chain $\Gamma$ starts in the first upper diagonal block of size $d^*$ and must end in the last upper diagonal block. In other words, at each step, the hidden Markov chain goes from the $k^{th}$ block to the $(k+1)^{th}$   block using the transition $i,j$ which corresponds to a probability law belonging to $\Rds{k}$.

\subsubsection{Trivariate Example}

\Comment{
\newcommand{\hsp}{\hspace*{0.15 cm}}
\newcommand{\mvgraph}[5]{\includegraphics[width=2.5cm]{V3/mpdf_#1-#2_#3_#4_kern#5}}
\newcommand{\mvgraphline}[2]{%
\mvgraph{#1}{#2}{0,1}{1}{1} \mvgraph{#1}{#2}{0,8}{1}{1} 
 \vline%
\hsp \mvgraph{#1}{#2}{0,1}{2}{1}  \mvgraph{#1}{#2}{0,8}{2}{1} 
 \vline%
  \hsp \mvgraph{#1}{#2}{0,1}{1}{2} \mvgraph{#1}{#2}{0,8}{1}{2}}
\begin{figure*}
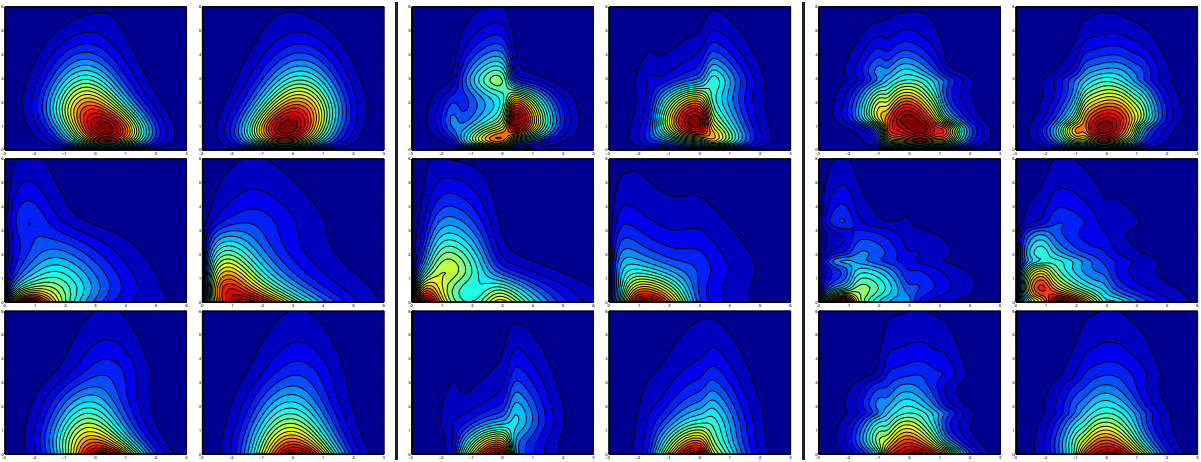


\begin{center}
\mvgraphline{1}{2}\\
\mvgraphline{2}{3}\\
\mvgraphline{1}{3}
\end{center}
\caption{ \label{fig:trivariate:1} Bivariate probability distributions of $\X$ (Left) and $\Y$ (Middle) and $\ZZ$(Right), for Pairs 1-2 (top), Pairs 2-3 (middle), Pairs 1-3 (bottom).  For each vector: Left: $p=0.1$. Right:  $p=0.8$.  
}
\end{figure*}

\newcommand{\uvgraph}[3]{\includegraphics[width=2.5cm]{V3/updf#1_#2_kern#3}}
\newcommand{\uvgraphline}[1]{\uvgraph{#1}{1}{1} \hspace{0.1 cm} \vline \hspace{0.1cm} \uvgraph{#1}{2}{1} \hspace{0.1 cm} \vline \hspace{0.1cm} \uvgraph{#1}{1}{2}  }
\begin{figure}
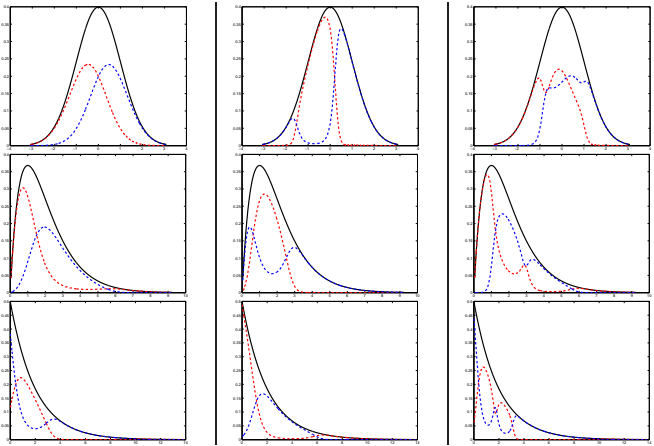


\begin{center}
\uvgraphline{1}\\
\uvgraphline{2}\\
\uvgraphline{3}
\end{center}
\caption{ \label{fig:trivariate:univ} Marginal distributions $\Prob_k$ (solid lines) for $\X$ (left), $\Y$ (middle) and $\ZZ$ (right) and the designed  $c_{i,j,k} \MP_{i,j}$ (dashed lines); for Components $X_1$, $Y_1$, $Z_1$ (top); Components $X_2$, $Y_2$, $Z_2$ (middle); Component $X_3$, $Y_3$, $Z_3$ (bottom).}
\end{figure}
}

Let us now develop the construction of two different trivariate random vectors $\X = (X_1,X_2,X_3)$ and $\Y =(Y_1, Y_2,Y_3)$, with marginal distributions set to a Gaussian distribution $\Normal{0}{1}$, for  $X_1$ and $Y_1$, to a Gamma distribution, with shape parameter $\alpha = 2$ and scale parameter $\beta = 1$ for $X_2$ and $Y_2$, and to a Gamma distribution with $\alpha=1$ and $\beta=2$ for $X_3$ and $Y_3$. 
To illustrate the potential of the tool,  $\X$ and $\Y$ also have the same correlations, but different joint distributions.  

First,  $d^*=2$, $\Estl=\alpha_0 I_d + \alpha_1 J_d$ and $\A^*_{i,j}=(1/d^*)$ are selected. 
Second,  to control correlations, the moment matrix $\Mqs{k}{1}$ is set, for both vectors, to ($k =1, 2 ,3$):
 \begin{equation}
\Mqs{k}{1}=\Estl \otimes
 \begin{pmatrix}
 m_{k,1}& m_{k,1}\\
 m_{k,2}& m_{k, 2}
 \end{pmatrix},
 \end{equation}
with the constraints:
\[m_{k,1}+m_{k,2}= 2 \Esp{X_k}.\]
Defining $\Delta_k= m_{k,1}-m_{k,2}$, the covariance reads
\[ \Covar{X_1}{X_2}=\Covar{Y_1}{Y_2}= (1-2\alpha_0) \Delta_1 \Delta_2, \] 
\[ \Covar{X_2}{X_3}=\Covar{Y_2}{Y_3}= (1-2\alpha_0) \Delta_2 \Delta_3, \] 
\[ \Covar{X_1}{X_3}=\Covar{Y_1}{Y_3}= (1-2\alpha_0)^2 \Delta_1 \Delta_3. \] 
Therefore, the covariance for any two consecutive components depends linearly on $\alpha_0$, and, when $\Delta_k \Delta_l>0$
is maximum for $\alpha_0=1$, vanishes at $\alpha_0=0.5$ and is minimum for $\alpha_0=0$.

The two trivariate joint distributions can now be made different via their moment matrices, $ \Mqs{k}{q}$, of order $q=2$, which 
for $\X$ is set to: 
 \begin{equation}
 \begin{aligned}
&\Mqs{1}{2}= \Estl \otimes 
 \begin{pmatrix}
 1 & 1\\
 1 & 1
 \end{pmatrix}, \,
  \Mqs{3}{2}= \Estl \otimes
 \begin{pmatrix}
 4.5& 4.5\\
 11.5 & 11.5\\ 
 \end{pmatrix}, \\
&\Mqs{2}{2}=  \Estl \otimes
 \begin{pmatrix}
 4.5& 4.5\\
 7.5 & 7.5\\ 
 \end{pmatrix}; \\
 \end{aligned}
 \end{equation}
and for $\Y$ to:
\begin{equation}
 \begin{aligned}
&\Mqs{1}{2}=  \Estl \otimes
 \begin{pmatrix}
 0.5 & 0.5\\
 1.5 & 1.5
 \end{pmatrix}, \quad
 \Mqs{3}{2}=  \Estl \otimes
 \begin{pmatrix}
  8& 8\\
  8 & 8
 \end{pmatrix}. \\
&\Mqs{2}{2}=  \Estl \otimes
 \begin{pmatrix}
  2.75& 2.75\\
 9.25 & 9.25
 \end{pmatrix}. 
 \end{aligned}
 \end{equation}

To construct the ${\MPs{k}}_{i,j}$ from the procedure developed in Section~\ref{sec:invprob}, a Gaussian kernel $K_{m,\sigma}(x)= \exp(-(x-m)^2/(2\sigma^2))$ is chosen. The resulting probability matrices $\MPs{k}$ have only $2$ distinct components. Fig.~\ref{fig:trivariate:univ} illustrates how the weighted density $c_{i,j,k}$ are combined in order to obtain the marginal distribution $\Prob_k$ for $\X$ (left) and $\Y$ (middle). 

Furthermore,  changing the Kernel $K$ parametrizing  $\MPs{k}_{i,j}$, also constitutes an efficient way to further vary the joint distributions, hence introducing further versatility in the procedure. 
For example,  $K$ could be chosen as the Gamma distribution family, or the union of the Gaussian and Gamma families.
To illustrate this, let us now construct a third trivariate random vector $\ZZ$, sharing the same marginals and covariance $\X$ and $\Y$ (it actually shares exactly the same matrices $\Mqs{k}{1}$ and $\Mqs{k}{2}$ as those of $\X$), though obtained from Kernel $K_{m,\sigma}(x)= (0.1+((x-m)/\sigma)^2) \exp(-((x-m)/\sigma)^2) $.

Bivariate partials distributions (rather than trivariate joint distributions, for clarity) are shown in
Fig.~\ref{fig:trivariate:1} for the pairs $(X_1,X_2)$ (top row), $(X_2,X_3)$ (middle row) and $(X_1,X_3$) (bottom row)  for $\X$ (left block), $\Y$ (middle block) and $\ZZ$ (right block), for negative (left column in each block) and positive (right column) correlations. 
These plots clearly show that, for any two pairs, the bivariate distributions (hence a fortiori the joint trivariate distributions) are different for the three vectors, though their components have same univariate marginals and same correlations. 
Furthermore, Fig.~\ref{fig:trivariate:univ} shows how the targeted marginal distributions $\Prob_k$ are obtained by summation of the $c_{i,j,k} \MPs{k}_{i,j}$ which were produced by the algorithmic procedure described in Section \ref{sec:invprob}. 
Note that with the chosen setting of the proposed examples, the summation has been a priori limited (for simplicity) to only $2$ different $\MPs{k}_{i,j}$, while $d^* \times d^* = 4 $ distinct distributions would have been available in general. 
Interestingly, comparisons for a given component (i.e., for a given row) amongst the three vectors illustrates that a same marginal distribution $\Prob_k$ is attained from different summation of the $c_{i,j,k} \MPs{k}_{i,j}$ for the three vectors, which constitutes a signature of the fact that the joint distributions of $\X$, $\Y$ and $\ZZ$ are different.

Here, the example was chosen trivariate, as a trade-off between ease of exposition (3 partial distributions of pairs of components remain easy to display) and demonstration of the potential and richness of the tool. 
Multivariate examples are however just as easy to construct.

\section{Conclusion and Perspectives}

A general framework for the theoretical definition and the practical numerical synthesis of random vectors and random time series, with a priori prescribed statistical properties, has been fully worked out, based on a matrix product formalism, and inspired from out-of-equilibrium statistical physics models.
Its ability to shape jointly marginal distributions and the dependence structure has been studied both theoretically and  practically. 
Pedagogical examples illustrated the versatility and richness of the procedure in actually attaining targeted properties. 
Remapping this matrix product framework onto that of Hidden Markov Models enabled us to devise an efficient practical numerical synthesis algorithm.
Also, a design procedure enabling to tune the elements of the models so as to reach desired targets has been obtained.
Both for design and numerical synthesis, {\sc Matlab} implementation of the described procedures are available upon request. 

Comparisons of the proposed approach to other numerical synthesis frameworks in terms of potential, versatility, efficiency, precision and implementation are under current investigations but are beyond the scope of the present contribution.
Further, the benefits of using other matrices $\A$ and $\E$ will be explored.
Moreover, having maintained the writing of the multivariate distributions as a product, as is the case for independent components, leads to possible computations of the distribution of the maximum $W= \max X_i$ or sum $S=\sum X_i$, of the components of $\X$. 
Such results are of premier importance for the use of such models in statistical physics applications as well as in signal processing for problems involving statistical properties of extremes or time-averages as ensemble average estimators. 
This is being investigated.
To finish with, the potential use of this synthesis tool to generate independent copies of sets of hyper-parameters in Monte Carlo Markov Chain numerical schemes constitutes a natural track to investigate.

\bibliographystyle{IEEEbib}
\bibliography{ProbaMat,ASEP,angeletti}

\end{document}